\documentclass{aa}
\usepackage{txfonts}
\usepackage{graphicx}
\usepackage{epsfig}
\usepackage{textcomp}

\newcommand{\beq} {\begin{equation}}
\newcommand{\enq} {\end{equation}}
\newcommand {\chisq}{\chi^{2}}
\def\lsim{\, \lower2truept\hbox{${< \atop\hbox{\raise4truept\hbox{$\sim$}}}$}\,}
\def\gsim{\, \lower2truept\hbox{${> \atop\hbox{\raise4truept\hbox{$\sim$}}}$}\,}
   
\begin{document}

\authorrunning{B.Cappellini et al. }
\titlerunning {Optimized in-flight absolute calibration for CMB surveys}  
\title{Optimized in-flight absolute calibration for extended CMB surveys}

\author{ 
   B. Cappellini \inst{1}, D. Maino \inst{1,2}, G. Albetti \inst{2},
   P. Platania \inst{1}, R. Paladini \inst{3}, 
   A. Mennella \inst{4}, M. Bersanelli \inst{1,4}
         }

\institute{
   Universit\`a degli Studi di Milano, via Celoria 16, I-20133 Milano, Italy
\and
   INAF-Osservatorio Astronomico, via G. B. Tiepolo 11, I-34131 Trieste, Italy 
\and
   SISSA/ISAS, Astrophysics Sector, via Beirut 4, I-34014 Trieste, Italy
\and
   IASF, CNR, via Bassini 15, I-20133 Milano, Italy
}

\offprints{Benedetta.Cappellini@mi.infn.it}
 
\date{Received October 8, 2002 / Accepted June 4, 2003}

\abstract{
Accurate measurements of the Cosmic Microwave Background (CMB) anisotropy 
call for high precision and reliability of the in-flight calibration.
For extended surveys the CMB dipole provides an excellent calibration source
at frequencies lower than $\sim$ 200 GHz; however poorly known foreground emissions, 
such as diffuse galactic components, complicate the signal and introduce 
a systematic error in the calibration. We show that  
introducing a weight function that takes into account the uncertainty in the 
a priori knowledge of the sky, allows us to substantially improve 
the calibration accuracy with respect to methods involving galactic 
latitude cuts. 
This new method is tested for {\sc Planck-LFI} radiometers at 30 and 100 GHz. 
On short time scales (less than 1 day) the absolute calibration of each channel
can be recovered with an overall 1-2\% accuracy.
We also consider the effect of CMB anisotropy itself on the calibration, and 
find that knowledge of the CMB pattern on large scales is needed to keep the 
short-time scale calibration accuracy within 1\%.
}

\maketitle



\section{Introduction}

Accurate determination of the CMB 
angular power spectrum can set powerful constraints on 
cosmological models and allows the determination of main cosmological 
parameters with high precision.
Recently, a number of ground-based and balloon-borne experiments
(\cite{boomerang}, \cite{maxima},
\cite{DASI}, \cite{CBI}, \cite{archeops})
have obtained remarkable evidence for the presence of acoustic 
peaks in the power spectrum, extending the pioneering measurements of 
$COBE$-DMR (\cite{COBE}) to sub-degree scales, and opening up the path 
to the next generation of space-based precision 
experiments (\cite{review}).
In fact, it has become clear (e.g. \cite{onlysat}) that only
space-based observations can provide the unique combination of 
environmental stability, freedom from systematic effects and avoidance of 
ground and atmospheric radiation needed for a high-resolution full sky survey.
The WMAP satellite by NASA was launched on June 2001 and the first-year data 
are now available.
The {\sc Planck} satellite is an ESA mission planned 
to be launched in 2007.

One of the most significant systematic uncertainties affecting current CMB 
anisotropy experiments is the instrument calibration, typically performed 
by observing celestial sources with known intensity, such as 
planets (usually Jupiter), strong radio sources 
(such as Cas A, Carina Nebula, Taurus A) or the CMB dipole. 
The accuracy is limited by the knowledge of the 
absolute flux ($\sim 5\%$ for planets, $\geq 8-10\%$ for radio sources and 
$\sim 0.4\%$ for CMB dipole), 
and also by the limited sky coverage in the case of the dipole.
The WMAP and {\sc Planck} surveys require great calibration 
accuracy, proportional to their instrument sensitivities.

To evaluate the accuracy actually achievable, all possible effects 
(both instrumental and astrophysical) 
impacting the calibration procedure have to be properly evaluated. 
In this work we focus on astrophysical effects by studying the presence
of different sky emissions
contaminating the main CMB dipole calibrator.
In particular, we propose a method to evaluate the accuracy one can 
expect for absolute calibration on short time scales, 
by using the available information on the
sky microwave emission.
With this method, that can be applied to any precision CMB 
measurement on large sky areas, it will be possible to 
further improve the calibration 
accuracy any time more precise sky observations become available
by simply reapplying the method we describe.
In a forthcoming paper we will analyse
instrumental effects (e.g., $1/f$ noise, pointing uncertainties, 
thermal fluctuations, gain drifts)  on the calibration performance.

In the present work we outline the proposed method and test the 
calibration technique for the {\sc Planck-LFI} receivers.

The structure of the paper is as follows:
in Sect. \ref{sect:concept} we briefly introduce the concept of calibration;
in Sect. \ref{sect:foregrounds} dipole contaminations (foregrounds) considered 
in the analysis are described.
Sect. \ref{sect:methodxerr} shows the method to compute statistical and systematic 
errors in the calibration performance, and describes the different 
calibration techniques tested by means of simulations.
Sects. \ref{sect:lfi-calib} and \ref{sect:LFI-results} are devoted to illustrating 
simulations for {\sc Planck-LFI} and consequent results.
In Sect. \ref{sect:cmb} the impact of CMB anisotropies on calibration accuracy
is studied.
The main conclusions and future works are discussed in Sect. \ref{sect:conclusions}.


\section{Calibration concept}
\label{sect:concept}

Calibration is the conversion of the output signal of each detector 
channel (e.g. telemetry units for a space experiment) to physical units 
(i.e. antenna temperature\footnote{The antenna 
temperature ($T_A$) is proportional to the detector power $P$ per bandwidth
$\Delta\nu$: $T_A=P/(k\Delta\nu)$; it 
depends on both source and receiver properties. The relation between 
antenna and thermodynamic temperature is $T_A=\frac{x}{e^x-1}T$ for 
intensities and $\Delta T_A= \Delta T\frac{x^2\, e^x}{(e^x - 1)^2}$
for temperature fluctuations; $x=h\nu/kT$, 
$h$ is the Planck constant, $k$ is the Boltzmann constant and $\nu$ is the 
observing frequency.}). 
In general, if the instrument response can be considered linear, then
\beq
T_A=G_0\, V + T_{\rm offset} \, ,
\enq
where $V$ is the receiver voltage output corresponding to the 
sky antenna temperature $T_A$, $T_{\rm offset}$ is an instrumental 
offset and $G_0$ is the calibration factor, constant to first order. 
In practice, $G_0$ can vary due to instrumental effects, 
e.g. amplifiers gain or thermal instabilities.

For a differential measurement, as in the case of anisotropy experiments,
the calibration is determined using pairs of sources:
\beq
T_2 - T_1 = G_0 (V_2-V_1) \,\, , 
\label{eq:Gdef}
\enq
where $T_2 - T_1=\Delta T_A$ is the antenna temperature difference of 
sources in the sky (often a well-known bright point source and the 
sky background)
and $V_2-V_1=\Delta V$ is the corresponding radiometer output. 

In order to measure the value of cosmological parameters with great 
precision, calibration
must be performed at $\sim 1\%$ overall  accuracy (a simple estimate 
of calibration requirement is shown in Appendix \ref{app:requirements}). 

The error in the determination of $G_0$ depends on the a priori 
uncertainty on the 
calibration sources temperatures, $\sigma_{\Delta T_A}$, and on the 
intrinsic detector noise, $\sigma_{\Delta V}$ (\cite{oldcalib}):
\beq
  \frac{\sigma_{G_0}}{G_0} \simeq \frac{\sqrt{\sigma_{\Delta 
T_A}^2+(G_0\sigma_{\Delta
  V})^2}}{\Delta T_A}\, ;
  \label{eq:sigmagoverg}
\enq
in this expression $\sigma_{G_0}$ accounts for both the statistical
error - intrinsic detector noise - and the systematic error
-the uncertainty
on the temperatures of the calibration sources-.
Eq.(\ref{eq:sigmagoverg}) shows that a better calibration is performed 
using higher ${\Delta T_A}$ (as long as the corresponding amplitudes do
not exceed the linear range of detectors). 

Since Eq. \ref{eq:sigmagoverg} is an estimate of the accuracy 
on $G_0$ using only a pair of points in the sky, and since
in extended surveys much more sky pixels are observed, 
the best value of the calibration constant is determined by 
fitting the distributions of $\Delta T_A$ and $\Delta V$, i.e. by
minimizing the one parameter $\chisq$ function:
\beq
\chisq(g)=\sum_k {\left[ \frac{\Delta V_k-\Delta T^{\rm cal}_k/g}
                           {\sigma_{\Delta V_k}} \right]^2} ~ ,
\label{eq:chi2}
\enq
where the index $k$ refers to the pixel pairs available for calibration.
In the following we indicate with $G$ the value of $g$ that minimizes 
$\chisq(g)$, while $G_0$ is the true value of the calibration factor, 
i.e. the value we need to recover; 
results will be expressed in terms of $G/G_0$.

\subsection{The CMB Dipole}
An observer in motion with velocity $\vec{\beta}=\vec{v}/c$ relative 
to the Planckian CMB field sees a dipole pattern: an angular 
distribution of the temperature given by
\begin{eqnarray}
T(\vartheta) & = & T_0 \frac{\sqrt{1-\beta^2}}{1-\beta {\rm cos}(\vartheta)} =
                  \nonumber \\
             & = & T_0+T_0 \beta {\rm cos}(\vartheta) + O(\beta^2) 
                     \nonumber ~ ,
\end{eqnarray}
where $T_0$ is the isotropic CMB temperature and $\vartheta$ is the angle 
between the direction of observation and the direction of $\vec{\beta}.$
The CMB dipole has been accurately measured by the $COBE$-FIRAS instrument 
(\cite{dipole}) with 
amplitude $\Delta T_{DIP}=T_0 \beta=3.372 \pm 0.014$ mK 
(i.e. $\vec v = 371 \pm 1$ km\,s$^{-1}$)
in the direction $(l,b)=(264.14^\circ \pm 0.15,48.26^\circ \pm 0.15);$
the overall error is $\simeq 0.4\%$.

The dipole is an ideal calibration source for CMB anisotropy experiments
covering a large sky area. Its amplitude is adequate
(not too strong to cause non-linear effects) and it allows a continuous 
calibration with no reduction of the observation time since it 
always enters the antenna's field of view.


\section{Foregrounds emission}
\label{sect:foregrounds}

Besides the CMB dipole, different components contribute to the radio-microwave 
brightness of the sky.
In this context, we are interested in these emissions (foregrounds) as 
``contaminations" of the prime calibrator, the CMB dipole.
In fact, these emissions are far less precisely known than the dipole
and so they represent a drawback for calibration, since they induce 
large errors in the determination of the $G$ factor.
In our analysis we only considered diffuse 
components: synchrotron, free free and interstellar dust emission.
We did not include additional diffuse components, 
such as emission from spinning dust grains (\cite{grains}), 
since a spatial and spectral full-sky template is not 
yet available for such components.
We also did not include point sources (e.g. galactic {\sc Hii} regions or
supernova remnants) since they fill only a very small fraction 
of the sky pixels.
To prove their low impact we considered as an example the effect of {\sc Hii} 
regions; results will be discussed in Sect. \ref{sect:LFI-results}. 

Amplitude and spatial distribution of the considered components have been
modeled (from available extended surveys) to produce synthetic 
sky maps at frequencies typical of CMB anisotropy experiments.   
We also estimated uncertainties on the intensity of each sky pixel for every 
foreground component.
Maps are represented with the 
HEALPix\footnote{\tt http://www.eso.org/science/healpix} 
pixelization scheme (\cite{healpix}).

In this way we are able to produce three sky maps
at each frequency of interest: 
a ``calibrator'' sky, i.e. the best knowledge of the
sky we can infer today from available data:
\beq
     T^{\rm cal}(\alpha,\delta) = \sum_i{T^{\rm cal}_i(\alpha,\delta)} ~,
\label{eq:tcal}
\enq 
an ``observed'' sky, 
which represents a deviation from $T_{cal}(\alpha,\delta)$
according to the uncertainties 
on the intensity of the considered components:
\beq
     T^{\rm obs}(\alpha,\delta) = \sum_i{T^{\rm obs}_i(\alpha,\delta)} = 
                                  \sum_i(T^{\rm cal}_i + \sigma_{T_i})(\alpha,\delta)
\label{eq:tobs}
\enq 
and a full-sky map of 
errors $\sigma_T(\alpha,\delta)$ obtained adding in quadrature 
the error for each component:

\beq
     \sigma_T(\alpha,\delta)=\sqrt{\sum_i{\sigma^2_{T_i}(\alpha,\delta)}} ~ .
\label{eq:terr}
\enq
Note again that the nature of $\sigma_T$ is systematic.

The sum in Eq. \ref{eq:tcal}, \ref{eq:tobs} and \ref{eq:terr} 
is extended to the components considered in the simulation.
The terms on the right side of Eq. \ref{eq:tcal} and 
\ref{eq:terr} are evaluated as follows.

The systematic uncertainty (1 $\sigma$) on the dipole is estimated 
from the COBE uncertainty on its amplitude and direction. 
To be conservative we considered the worst
case for calibration (see Sect. \ref{sect:LFI-results-dipole}): 
when considering dipole, the ``observed'' component is  a 
``stretched'' dipole with $\vec v =370$ km\,s$^{-1}$,  
$l = 263.99^{\circ}$ and $b = 48.41^{\circ}$.

To estimate synchrotron emission we used the Haslam et al. (1982) 
all-sky survey at $\nu_0=408$ MHz, properly rescaled 
using a spectral indices $\beta_i$ map:
\beq
T_i(\nu)=T_i(\nu_0)\left( \frac{\nu}{\nu_0} \right)^{\beta_i} \, .
\enq
The spatial distribution of the synchrotron spectral index $\beta_i$ 
was obtained
using the 408 MHz, 1420 MHz (\cite{1420MHz}) and 2326 MHz 
(\cite{2326MHz}) maps. To derive errors 
we took into account the 10\% uncertainty of the original Haslam map 
plus an error on the spectral indexes determination; 
these errors values lie in the range [0.1,0.8]. 

Dust emission was estimated using the 100 $\mu$m intensity and 
100/240 $\mu$m flux ratio map by Schlegel et al. (1998), generated from 
IRAS and $COBE$/DIRBE data\footnote{Available at 
{\tt http://space.gsfc.nasa.gov/astro/cobe/
dirbe\_products.html}}. 
They also provide (\cite{dust_extrap}) tools for dust emission
extrapolation at CMB frequencies, using FIRAS data: we used their 
best model, a 2-components model (n.8 in their paper) with mean dust 
temperatures $<T_1>=9.4$ K and $<T_2>=16.2$ K, and spectral indices 
$\alpha_1=1.67$ and $\alpha_2=2.70$. 
The authors also estimate the 
errors on extrapolated maps to be $\sim$ 10\%.

Diffuse free-free emission is poorly known.
We only considered a free-free component spatially correlated with 
dust, with an intensity equal to 30\% of dust at 100 GHz 
(\cite{dezotti}), and rescaled in frequency using a power low relation:
\beq
T(\nu)=T(100)\left( \frac{\nu}{100} \right)^{\alpha_{ff}} \, \, ,
\enq
where $\nu$ is in GHz and $\alpha_{ff}=-2.1$, the typical Bremmstrahlung 
spectral index.
The estimated free-free emission error is 10\% on dust uncertainties 
plus 10\% on the correlation factor.

Numerical results for the systematic uncertainty on $G$ due to the presence 
of poorly known foregrounds, highly depend on these estimates; 
anyway, the ability of the weight function technique to 
reduce significantly this uncertainty was successfully tested 
with other scenarios for the ``errors'' maps, so that the proposed method 
could also be applied in more favourable situation, 
e.g. taking information from the recent WMAP mission maps of the 
microwave emission.


\section{Determination of statistical and systematic errors
         on calibration performance}
\label{sect:methodxerr}

To evaluate both the statistical and systematic errors on the recovered
gain factor, each complete simulation is split into 2 steps: 
\begin{enumerate}
\item We set to zero the $\sigma_T(\alpha,\delta)$ terms, so that 
   the same sky map is considered: 
   $T^{obs}(\alpha, \delta)=T^{cal}(\alpha, \delta)$; this is
   an ideal situation where we assume perfect knowledge of the observed signal.
   This first simulation allows us to evaluate the statistical component of 
   the error due to instrumental noise.
\item A sky map $T^{cal}(\alpha, \delta)$ different from $T^{obs}(\alpha, \delta)$
   is considered (Eq. \ref{eq:tcal} and \ref{eq:tobs}).
   In this case both the uncertainties (astrophysical and
   instrumental noise) affect the simulation: 
   the difference between the result of this second
   simulation and the first one gives the systematic component 
   of the error.
\end{enumerate}
Of course in both simulations the same noise realization has to be 
used.

\vspace*{0.2cm}
First of all we analyse the ideal case where only the CMB dipole is considered; 
this allows us to evaluate the systematic error due to CMB dipole 
uncertainties.
In this simple case, the Eq. \ref{eq:tcal} and \ref{eq:tobs}
are simplified as follows:
$
T^{obs} = T^{dip}
$ ,
$
T^{cal} = T^{dip}_{stretched}
$ (see Sect. \ref{sect:foregrounds}).

Secondly, to consider a more realistic situation, we introduce 
galactic emissions: we analyse the impact on $G$ of foregrounds 
uncertainties only, i.e. in these simulations the same dipole 
is considered both in the observed and in the calibrator skies, 
to fully analyse one type of problem at each time.

A way to handle the problem of large errors due to foreground 
emissions is to cut the galactic plane in the data analysis;
this was for example the solution adopted by the $COBE$ team in the analysis 
of DMR data (\cite{DMRcal}). This technique is limited by the presence 
of emission at high galactic latitudes, and is not preferable since it 
forces one not to consider a portion of data.

\begin{figure}[htb]
\mbox{}
\centerline{\epsfig{figure=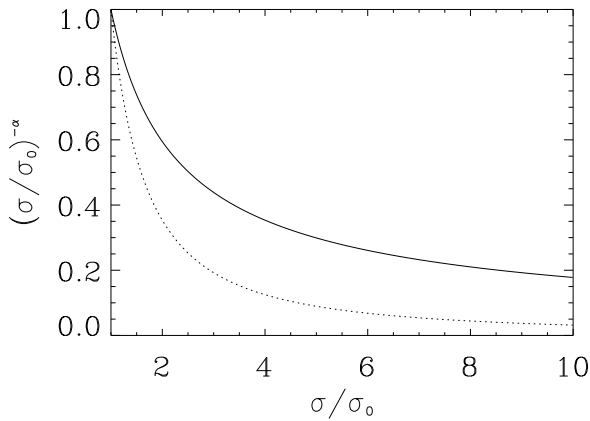, height=6.5cm, angle=0}}
\caption[]{Weight function for two fixed values of the $\alpha$ 
parameter: 0.75 (solid line) and 1.5 (dashed line).}
\label{fig:weight_function}
\end{figure}

In order to make maximum use of the whole data set, we looked for a
technique able to properly weight the signal intensity $T$
of each point of the sky depending on its 
uncertainty $\sigma_{T}$. This can be done introducing a suitable 
{\it weight function}.
Several functions were considered; based on our analysis we concentrated 
on the family of simple functions
\beq
W(\sigma_{T})=\frac{1}{(\sigma_{T}/\sigma_0)^\alpha}\, , 
                        \qquad \alpha\in\Re
\label{eq:weight_func}
\enq
where 
\beq
\sigma_0=\inf_i{(\sigma_{T_i})} \, .
\enq
For every value of $\alpha$ this is a decreasing function, with 
max$(W(\sigma_{T}))=W(\sigma_0)=1$
(Fig. \ref{fig:weight_function}). 

When considering pixel-pairs, we use $W(\sigma_{\Delta T})$
where
\beq
\sigma_{\Delta T} = \sqrt{ \sigma_{T_1}^2+\sigma_{T_2}^2 } \; ,
\enq
and
\beq
\sigma_0 = \inf_{i} (\sigma_{\Delta T_{i}}) \, .
\enq

Eq. (\ref{eq:chi2}) is then modified as follows:
\beq
\chi^2_W (g)=\sum_k { W(\sigma_{\Delta T_k}) 
                \left[ \frac{\Delta V_k-\Delta T^{\rm cal}_k/g}
                           {\sigma_{\Delta V_k}} \right]^2} \, . 
\label{eq:chi2_weight}
\enq

In this way the contribution of pixel-pairs with a poorly known 
temperature is reduced (at a rate controlled by $\alpha$), leading 
to a great improvement in the calibration accuracy.
The value of the $\alpha$ parameter to be used is chosen in order 
to minimize the overall uncertainty on calibration.

For a given experiment and scanning strategy, one can study the trend 
of both the statistical and systematic errors,
$\sigma_{stat}(\alpha)$ and $\sigma_{syst}(\alpha)$
respectively, computing the standard deviations of calculated errors 
over the whole mission lifetime as a function of $\alpha$.
One expects the weight technique to act in opposite ways on the two 
types of errors:
while the systematic error is a decreasing function of $\alpha$, the 
statistical error is an increasing one.
To optimize this procedure one has to search for the minimum of 
the function
\beq
f(\alpha)=[ \sigma^2_{\rm syst}(\alpha) + 
            \sigma^2_{\rm stat}(\alpha) ] \, ,
\label{eq:f_alpha}
\enq 
yielding the optimum parameter $\alpha$.


\section{Calibration for {\sc Planck-LFI} receivers} 
\label{sect:lfi-calib}

{\sc Planck} will observe CMB 
anisotropy and polarization with an unprecedent combination of sensitivity, 
sky coverage, frequency range and angular resolution.
It consists of two instruments sharing the focal plane
of an off-axis aplanatic 1.5 meter aperture telescope.
The Low Frequency Instrument ({\sc LFI}) covers the range 30-100 GHz with 
four observational channels centered at 30, 44, 70 and 100 GHz.
The High Frequency Instrument ({\sc HFI}) will observe the sky in six channels
between 100 and 857 GHz with bolometers detectors. The wide frequency 
coverage will allow the separation of CMB anisotropies from non 
cosmological signals, thanks to their different spectral behaviour.

We applied the absolute calibration procedures described in the 
previous section to the case of {\sc Planck-LFI} radiometers.
A first analysis of calibration feasibility was performed 
in the context of the COBRAS/SAMBA\footnote{COBRAS/SAMBA 
was the previous name of the {\sc Planck} satellite.} mission study
(\cite{oldcalib}).
The {\sc HFI} consortium has analysed the problem as well (\cite{HFIcalib}),
focusing on short time scale relative calibration, and long time scale 
($\geq$ 1 month) absolute calibration based on the CMB and Earth orbital 
motion dipoles.

During operations calibration must be performed with high 
accuracy and as frequently as possible, to control possible drifts on 
instrumental gain and offsets.
The monitoring of receivers response requires a relative 
calibration on short time scales, while, in principle, absolute calibration 
can be only performed at the end of the operations and recovered through the
whole mission lifetime thanks to the relative calibration.
For {\sc Planck-LFI} data, a relative calibration will be performed 
between the 1-hour circles thanks to the observational scanning strategy.
Regarding the absolute calibration,
for {\sc LFI} receivers
the main calibration source is the CMB dipole. 
Only on long time scales ($\gsim$ 3 months) an even better absolute calibration 
can be performed using the dipole modulation due to the 
satellite orbital motion around the Sun (which is known 
with high precision), as shown in Piat et al. (2002).
On the other hand, in the data analysis process it will be of high
interest to monitor the absolute calibration on short time scales (in
particular for inter-frequency quick-look analysis comparisons) for which
the dipole modulation cannot be used. The method described in this paper
allows us to optimise the absolute calibration strategy of a full-sky mission
at short time scales. In addition, this concept can be applied to balloon
or ground-based experiments with limited sky coverage, for which the
dipole modulation is not an effective calibrator.

To simulate {\sc LFI} receiver calibration, 
first we created simulated data streams\footnote{The
TOD (Time Ordered Data) used in this work was generated by the {\sc Planck} 
pipeline simulator of the Level-S of the DPC. 
Sky simulations, pointings and
other data are available for {\sc Planck} collaboration at
{\tt http://planck.mpa-garching.mpg.de/SimData.}}
considering the {\sc Planck} baseline scanning strategy. 
According to this strategy, the satellite orbits 
around the L2 Lagrangian point of
the Earth-Sun system; the spacecraft spins at 1 r.p.m. around its 
spin axis which is 
kept along the ecliptic plane and is repointed by $2.5'$ every hour
to keep the anti-solar direction; the telescope field of view is at
an angle of $85^\circ$ from the spin-axis direction. 

Following the guidelines described in Sect. \ref{sect:foregrounds}, 
we generate sky maps at {\sc Planck-LFI} frequencies (30-100 GHz range); 
these maps are then converted into antenna temperature,
and convolved with a gaussian beam with the Full Width Half Maximum 
(FWHM) of {\sc LFI} beams (i.e. 33$'$ at 30 GHz and 10$'$ at 100 GHz)
by the pipeline simulator. 
The code includes the main properties of
the {\sc Planck} payload (e.g. the boresight angle, the scanning 
strategy) and of the considered receiver (e.g. the beam location on the 
focal plane, its FWHM, noise properties). 

The scanning strategy is such that every detector observes the same
``ring" in the sky for an hour before repointing; in our work 
we average the 60 
1-minute observations of the same sky-ring, considering 1-hour 
data streams; we refer to these observations as ``circles".

The FWHM of the antenna divides every circle ($j$) in pixels ($i$).
The radiometer output is
\beq
V_{ij}=\frac{T^{\rm obs}_{ij}+n_{ij}}{G_0} \, ,
\label{eq:signal}
\enq
where $T^{\rm obs}$ is the observed sky, $n$ is a noise term and $G_0$ 
is the true value of the calibration constant. 
In the noise term we only considered white noise, i.e. a random Gaussian 
distribution with rms
\beq
\delta T_{rms}=\sqrt{2}\frac{T_{sys}+T^{\rm obs}}{\sqrt{\Delta\nu\cdot\tau}} \,\, ,
\label{eq:rms_noise}
\enq
where 
$T_{sys}$ is the system temperature of the receiver, 
$\Delta\nu$ is the bandwidth ($\Delta\nu/\nu\sim$ 20\%) and 
$\tau$ is the considered integration time;
the noise amplitude $\delta T_{rms}$ is quite constant during the 
mission. 
In this case, the instrumental uncertainty on output 
differences (see Eq. \ref{eq:sigmagoverg}) is:
\beq
\sigma_{\Delta V} \simeq \frac{\sqrt{\delta T_1^2+\delta T_2^2}}{G_0} \, ,
\enq
where $\delta T_1$ and $\delta T_2$ are the rms values 
in the two observed points of the sky.

By averaging 60 rings in one circle we assume that the noise of each 
ring is uncorrelated with the others, which is a reasonable assumption
for the low $1/f$ noise expected from the instrument (\cite{oof}).

\begin{figure}[htb]
\mbox{}
\centerline{\epsfig{figure=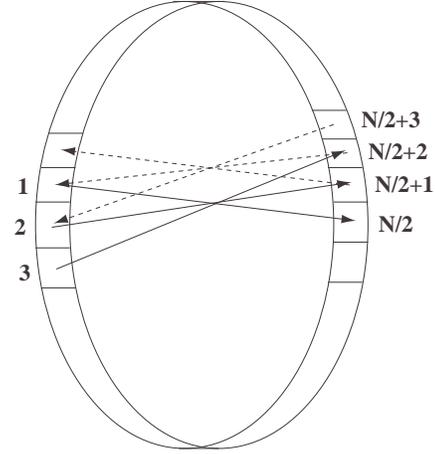,height=6cm, angle=0}}
\caption[]{Chosen pixel pairs.}
\label{fig:pairs}
\end{figure}

Secondly, we need to determine the pixel-pairs for calibration.
If the number of pixels in a circle is $N$, then $N-1$ is the number of independent 
pixel-pairs; we must choose them among the $N(N-1)/2$ possible pairs.
Different criteria for choosing the $N-1$ independent pixel pairs were 
considered: the results obtained showed no significant changes.
The simplest approach was then selected: pairs are formed with opposite 
pixels in a circle (Fig. \ref{fig:pairs}):

\begin{enumerate}
  \item $i \in [1, N/2+1] \rightarrow i'=i+N/2-1$
  \item $i \in [N/2+2, N] \rightarrow i'=i-N/2-1$
\end{enumerate}
In this way every pixel is considered twice; finally the 
pair with the lowest signal difference is 
not considered, thus leaving exactly $N-1$ pairs.

The $k$ parameter in the $\chisq$ function (Eq. \ref{eq:chi2}) becomes in this case 
$k=(m,j)$, where $m=(i,i')$ refers to pixel-pairs previously defined
and $j$ refers to the considered circle.

In our simulations the $G$ factor is computed over two different 
time-scales: 1 hour and 1 day. 
Both are ``natural" time scales for {\sc Planck}: the first defines 
the scan ``circles'' while the second is the time scale of
Earth-satellite communications. Thus, the sum in Eq. \ref{eq:chi2} is extended
to the $N-1$ 
independent pixel pairs of one circle in the first case (fixed $j$ value), 
while is extended to the $(N-1)\cdot 24$ circles covered in one day in the 
second case.


\section{Simulation results}
\label{sect:LFI-results}

We present the analysis carried out at 30 and 100 GHz. The first is the 
channel where foreground emission is the strongest while the 
second one has the highest noise level per single radiometer
(for 30 GHz radiometers $T_{sys} \simeq 10$K, 
 for 100 GHz $T_{sys} \simeq 45$K): 
they represent the worst 
case for systematic and statistical error behaviour respectively.
The other {\sc LFI} frequencies are in intermediate situations.

We applied the 2-step procedures described in Sect. \ref{sect:methodxerr}; 
as already noted, the same noise realization has to be used for both steps.
In the case of {\sc LFI} radiometers, since the noise rms value has a 
slight dependence 
on the observed sky temperature (see Eq. \ref{eq:rms_noise}), 
considering the same realization for both simulation introduces an error 
of a few \textperthousand ~ on signal terms (Eq. \ref{eq:signal}).

Similar results hold for all the radiometers at the same frequency.
At 30 GHz  we have considered the radiometer LFI-28  
with the beam position ($\theta_{B}=5.26^{\circ}$, $\phi_{B}=52.9^{\circ}$) 
in the focal plane and angular
resolution of $33'$ that divides every observed circle in 1950 
pixels\footnote{Note that the dimension of a pixel 
is equal to FWHM/3.}; 
at 100 GHz we used the radiometer LFI-01 with the beam position 
($\theta_{B}=2.93^{\circ}$, $\phi_{B}=0.0^{\circ}$) and the angular 
resolution of $10'$ that divides every observed circle in 6498
pixels$^6$.

\subsection{CMB Dipole}
\label{sect:LFI-results-dipole}
As described in Sect. \ref{sect:methodxerr}, we start by considering 
the CMB dipole only. This allows us to evaluate the systematic error 
due to CMB dipole uncertainties, and also provides a consistency test of our code.

\begin{figure}[htb]
\mbox{}
\centerline{\epsfig{figure=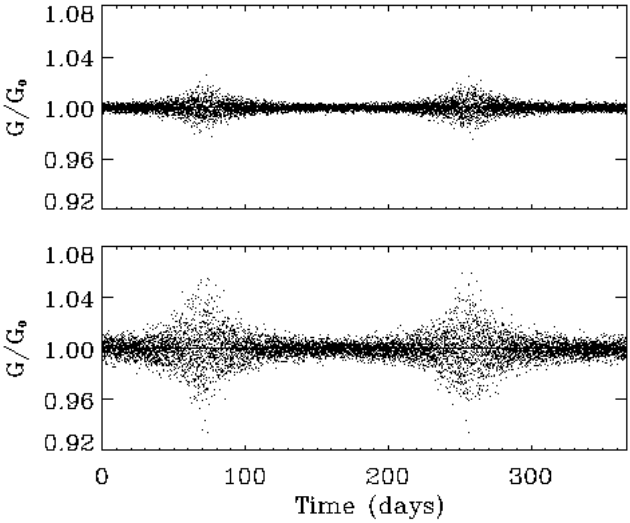, width=10cm, angle=0}}
\caption[]{Statistical error on the $G$ value calculated every hour at 30 GHz 
(upper panel) and 100 GHz (lower panel).}
\label{fig:stat_dip}
\end{figure}

\begin{figure}[htb]
\mbox{}
\centerline{\epsfig{figure=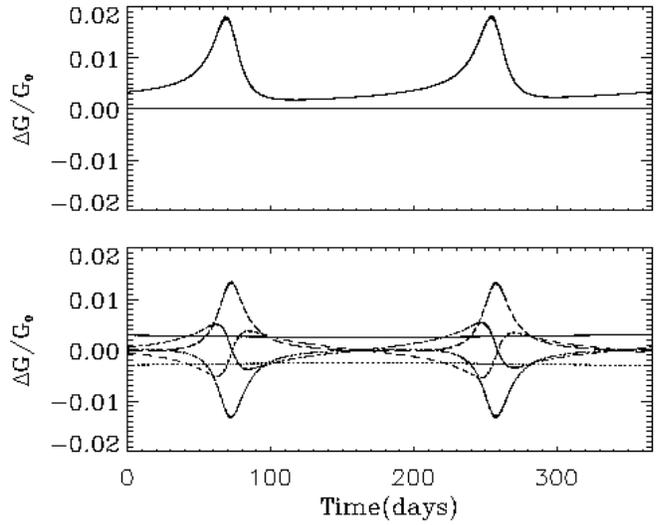, width=10cm, angle=0}}
\caption[]{Systematic error due to CMB dipole uncertainty (from the COBE 
measure) on the $G$ value calculated every hour, at 30 GHz 
and 100 GHZ (upper panel, plots almost perfectly overlap).
In the lower panel, ``partial'' systematic errors are shown 
(see text for more details); solid line: $\vec v = 370$ Km\,s$^{-1}$, 
dotted line: $\vec v = 372$ Km\,s$^{-1}$,
dashed line: $l=264.29^\circ$, dotted-dashed line: $l=263.99^\circ$, 
double dotted-dashed line: $b=48.11^\circ$, long dashed line: $b=48.41^\circ$.}
\label{fig:syst_dip}
\end{figure}

\begin{figure}[htb]
\hspace*{0.2cm}
\mbox{}
\centerline{\epsfig{figure=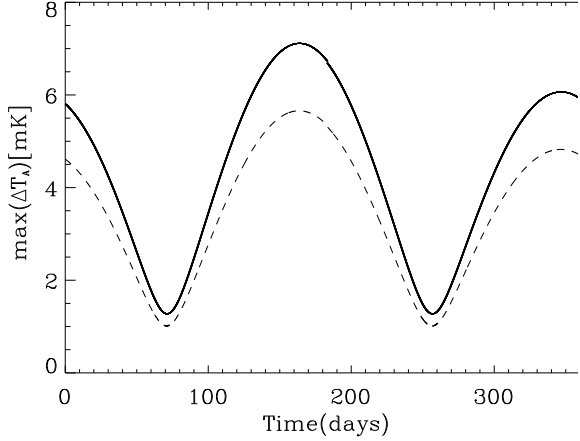, width=9.2cm, angle=0}}
\caption[]{Higher $\Delta T$ available for every hour-calibration at 30 GHz 
(solid line) and 100 GHz (dashed line).}
\label{fig:max_dip}
\end{figure}

Results on the 1-hour time scale are shown in Fig. \ref{fig:stat_dip} (statistical 
errors) and Fig. \ref{fig:syst_dip} (systematic errors);  
on such short time scales the statistical and the systematic errors are 
of the same order, $\lsim 1-2\%$. 
The typical trend of the statistical error is the effect of different 
$\Delta T_{dip}$ values ($\sigma_{G_0} \propto \Delta T_A^{-1} $, 
cfr. Eq. \ref{eq:sigmagoverg}) observed 
in the scan circles over the different periods of the mission time, 
due to the changing geometry of the field of view with respect to the 
dipole direction. The correlation is clear looking at Fig. 
\ref{fig:max_dip}, where the largest temperature difference of every circle is 
plotted.
The shape of the systematic error depends on the choice of $\vec v, l$ and $b$ in
the stretched dipole. In Fig. \ref{fig:syst_dip} (lower panel) the effect on $G$ 
when changing these parameters one at a time is shown. To be conservative we 
chose a combination (see Sect. \ref{sect:foregrounds}) of these parameters 
that maximize the dipole systematic error (Fig. \ref{fig:syst_dip}, upper panel).

As expected, the amplitude of the statistical and systematic errors are 
consistent with the estimate of detector sensitivities on the considered 
time scale ($\sim$ 0.2 mk Hz$^{-1/2}$ at 30 GHz and 
$\sim$ 0.6 mk Hz$^{-1/2}$ at 100 GHz)
and with the $COBE$-FIRAS uncertainties on dipole measurements.

\begin{figure}[htb]
\mbox{}
\centerline{\epsfig{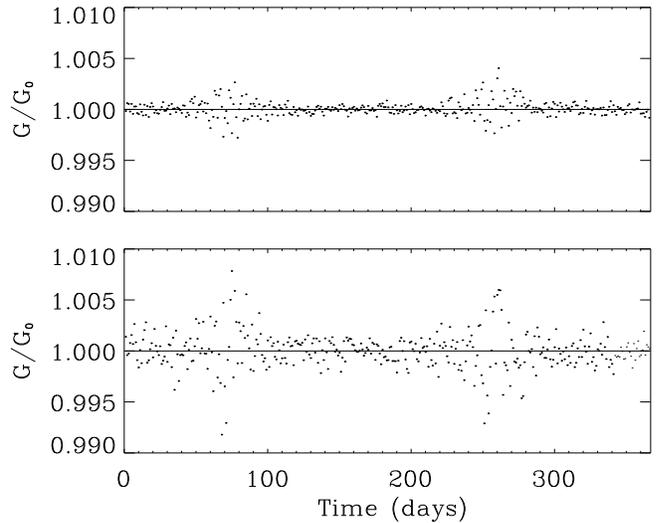}}
\caption[]{Statistical error on the $G$ value calculated every 24 hours at 30 GHz 
(upper panel) and 100 GHz (lower panel).}
\label{fig:stat_dip24}
\end{figure}

If calibration is performed every 24 hours, the statistical error 
(Fig. \ref{fig:stat_dip24}) shows the same trend over time but  
with much lower amplitude (cfr. Eq. \ref{eq:rms_noise}); 
the systematic error is time independent and indeed results over different 
time scales perfectly overlap (results with 24 hours as time-scale are thus not shown).
These time-trends of  statistical and systematic errors
always apply for a given calibration technique if
the only difference  is the time scale chosen 
to recover the gain factor.

\subsection{Foregrounds}
\begin{figure}[htb]
\hspace*{0.2cm}
\mbox{}
\centerline{\epsfig{figure=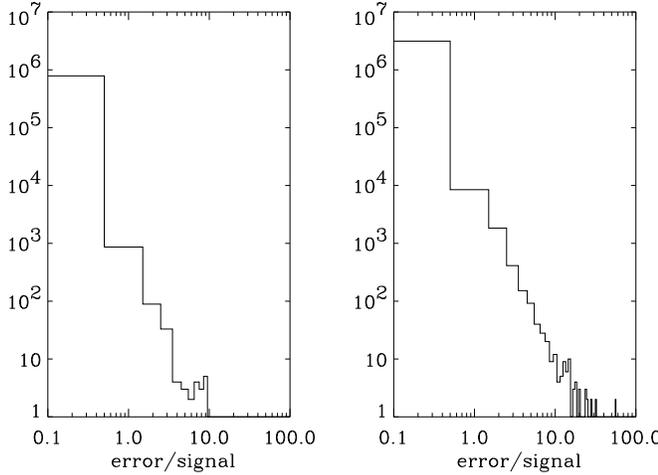, width=9.2cm, angle=0}}
\caption[]{Histograms for the ``error"/``observed'' maps of foregrounds
at 30 (left panel) and 100 (right panel) GHz.}
\label{fig:hist-gal/err}
\end{figure}

A more realistic situation is obtained introducing galactic emissions.
Histograms for the ``error''/``observed'' maps of foregrounds 
(Sect. \ref{sect:foregrounds})
at the considered frequencies are plotted in Fig. \ref{fig:hist-gal/err}.
In both cases, for less than 0.4\% of the sky pixels 
this value is greater than 1, while isolated pixels have 
very high ``error"/``observed "values.

\begin{figure}[h!]
\mbox{}
\centerline{\epsfig{figure=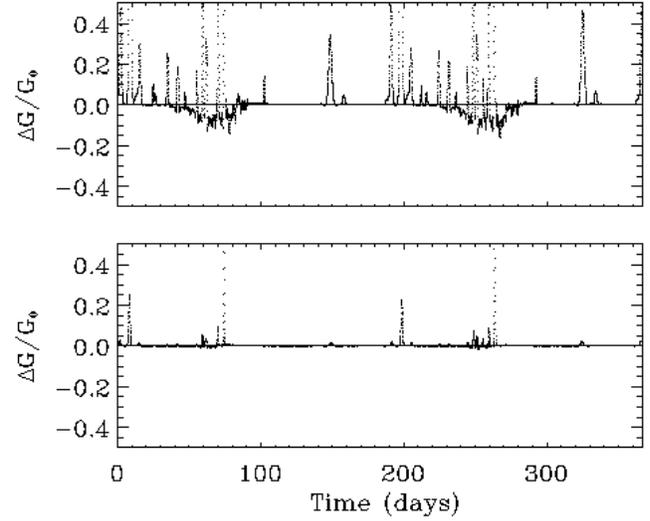, width=10cm, angle=0}}
\caption[]{Systematic error due to galaxy uncertainty on the $G$ value 
calculated every hour, at 30 GHz (upper panel) and 100 GHz (lower panel).}
\label{fig:syst_gal}
\end{figure}

Fig. \ref{fig:syst_gal} shows the systematic error induced by uncertainties 
on foregrounds components.
At both frequencies the errors are much larger than required (see Appendix
\ref{app:requirements}) and, as expected, the situation is far more 
problematic at 30 GHz. 

The presence of spikes in Fig. \ref{fig:syst_gal} means that very large errors are
concentrated in limited sets of scan circles.
In the following we give an explanation of the presence of such spikes.

As already mentioned in Sect. \ref{sect:foregrounds}, these simulations only 
consider diffuse emission components.
Anyway we investigated the impact of point sources considering 
galactic {\sc Hii} regions; we used a recent compilation of 1442 sources
(\cite{HIIregions}): they affect less than
0.5\% of the sky (\cite{HIInote}), contaminating $\sim  7\%$ of scan circles in very few pixels.
To be conservative, their emission was only included in the 
$T^{obs}(\alpha,\delta)$ map: we thus calibrated assuming no information at 
all on their presence. 
Simulations show there is no impact on the statistical error, while
the systematic error is increased (only in the 
contaminated scan circles): in practice other spikes similar to those
present in Fig. \ref{fig:syst_gal} appear in the $G$-plot.
Furthermore we run the calibration procedure excluding in the 
$\chisq$ function (Eq. \ref{eq:chi2}) the pixels 
with known {\sc Hii} regions; again the statistical error is not 
affected, but we found that most of the spikes in 
Fig. \ref{fig:syst_gal} disappear.
This result can be explained in a simple way:
spikes in the $G$ plots are due to the largest values in our 
estimated error map, which correspond -not a surprise-
to point source pixels, and to sources not completely removed in the 
maps of the the dust component.

These tests show that point sources do not interfere significantly with
the proposed calibration technique, 
as long as they can be removed with the aid of source catalogues.

On the other hand strong point sources, such as brightest {\sc Hii} regions 
and planets, can be very useful as a supplementary way to monitor the 
stability of the gain factor ({\it relative calibration}), 
since they are clearly 
recognizable in the received signal given their high intensity. 
Indeed, relative calibration only requires stable bright sources, 
and not a precise knowledge of their absolute intensity.
When possible, relative calibration on these sources can be performed over 
different time scales, such as:
\begin{itemize}
\item less than 1 hour, since they are observed by a given detector every 
      minute with a high signal to noise ratio;
\item six months, i.e. the time needed for a given detector to come back 
      to the same portion of the sky.
\end{itemize}

\subsubsection{Galactic cuts calibration technique}
As a further step, we simulated
the calibration technique with a $\pm 20^{\circ}$ galactic cut; 
higher galactic cuts do not permit calibration over long periods 
during the mission, 
due to an excessive decrease in the number of avilable pixel-pairs.

\begin{figure}[h!]
\mbox{}
\centerline{\epsfig{figure=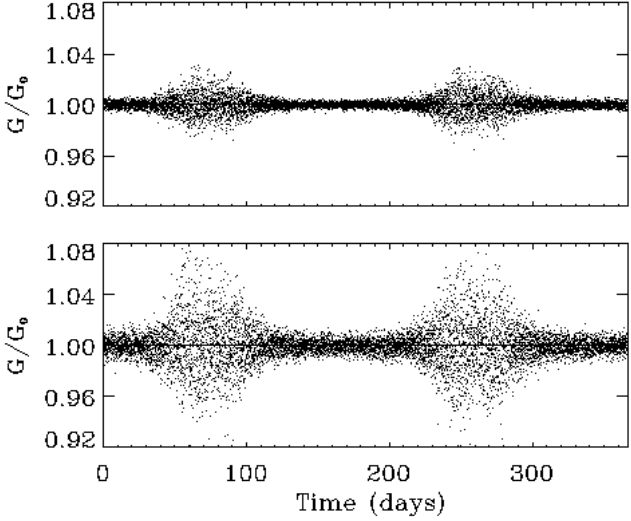, width=10cm, angle=0}}
\caption[]{Statistical error on the $G$ value calculated every hour,
with a galactic cut of $20^\circ$, 
at 30 GHz (upper panel) and 100 GHz (lower panel).}
\label{fig:stat_cuts}
\end{figure}

\begin{figure}[h!]
\mbox{}
\centerline{\epsfig{figure=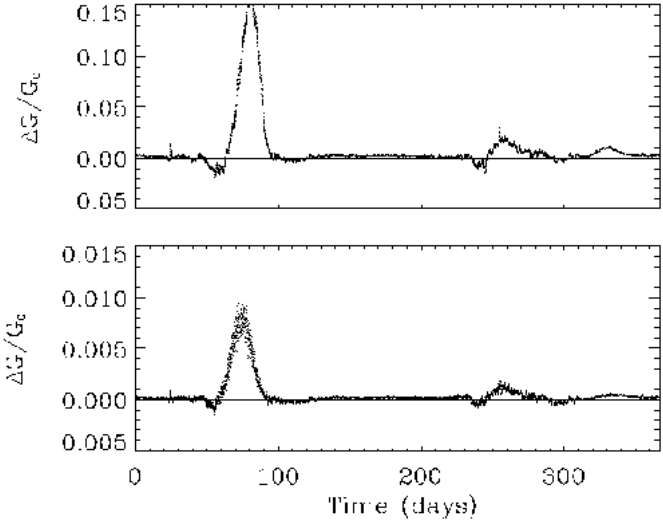, width=10cm, angle=0}}
\caption[]{Systematic error on the $G$ value calculated every hour,
with a galactic cut of $20^\circ$, 
at 30 GHz (upper panel) and 100 GHz (lower panel).
The spike in the results is the effect of Magellanic 
Clouds emission at high galactic latitudes.}
\label{fig:syst_cuts}
\end{figure}

We only show results with a 1-hour integration time, in Figs. \ref{fig:stat_cuts} 
and \ref{fig:syst_cuts}.

As expected, the reduction in the data-set increases the statistical error 
(Fig. \ref{fig:stat_cuts}, compare to Fig. \ref{fig:stat_dip}); on the other hand 
the systematic error is highly reduced as can be seen comparing 
Fig. \ref{fig:syst_cuts} to Fig. \ref{fig:syst_gal}.
Anyway this technique is limited by the presence of residual emission at
high galactic latitudes; in Fig. 
\ref{fig:syst_cuts} we can see the effect of the emission of 
Magellanic Clouds that are at $-30^{\circ}$ of galactic latitude. 
At 30 GHz their signals uncertainty produces an error of $\sim$15\% on $G$.

\subsubsection{The ``weight function" calibration technique}
\begin{figure}[h!]
\mbox{}
\centerline{\epsfig{figure=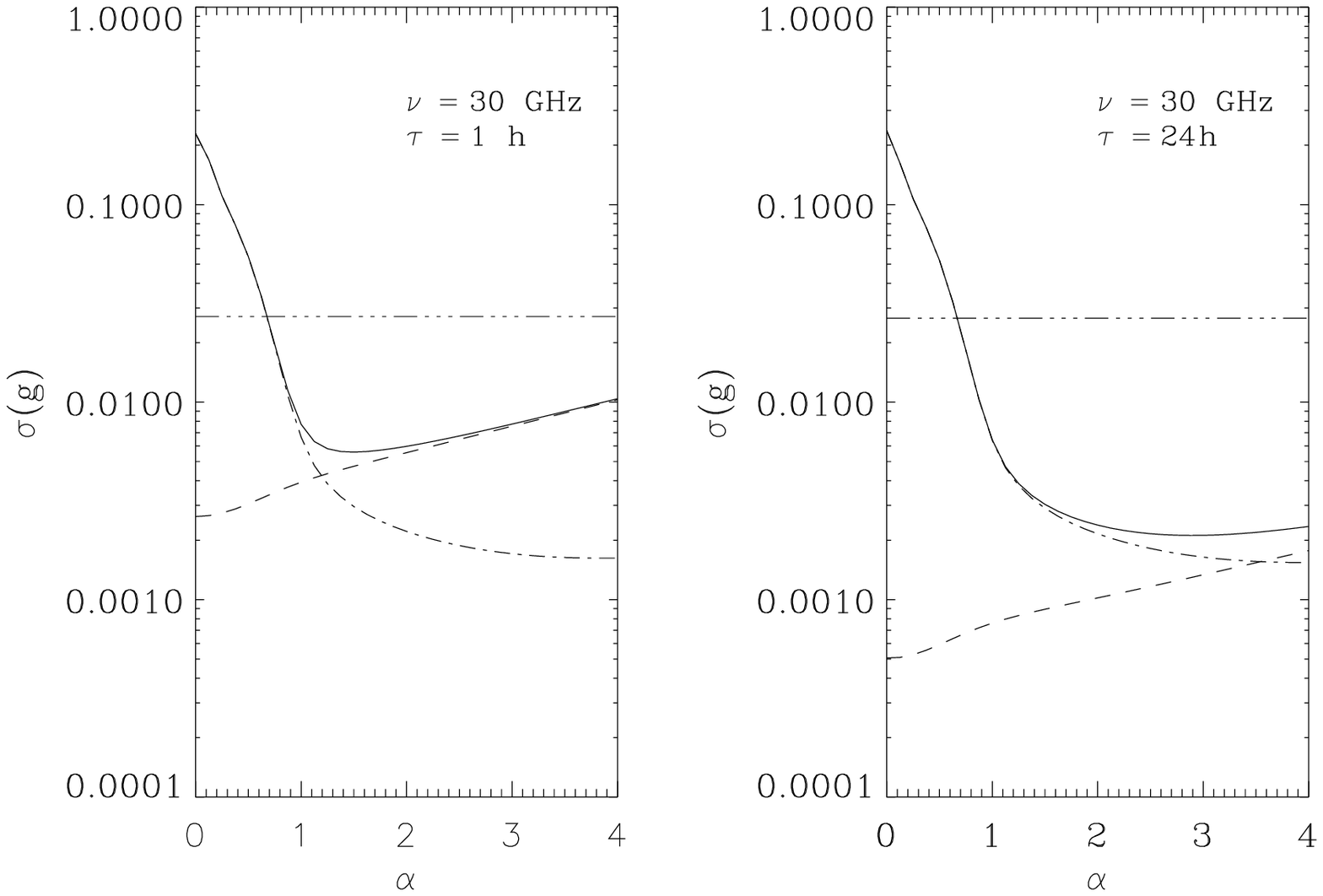, width=9.2cm, angle=0}}
\centerline{\epsfig{figure=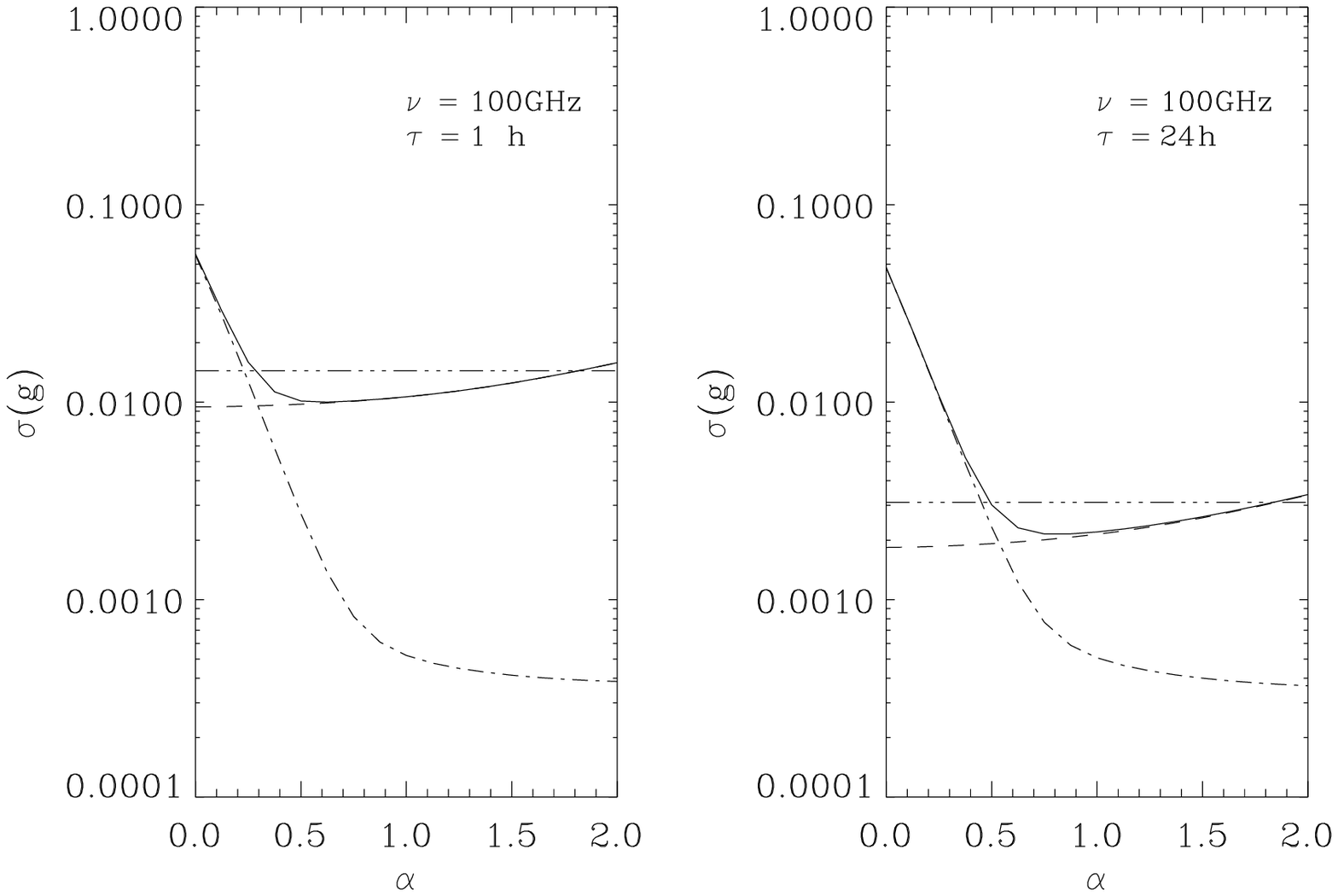, width=9.2cm, angle=0}}
\caption[]{Systematic (dot-dashed line) and statistical (dashed line) error 
behaviour as a function of the parameter $\alpha$ (defined in Eq. 
\ref{eq:weight_func}). The best $\alpha$ value is the one 
minimizing $f(\alpha)$ (solid line, see Eq. \ref{eq:f_alpha}). 
The horizontal line is the total error for the 
$20^\circ$ galactic cut technique. 
Panels on the left correspond to a 1 hour timescale calibration; 
those on the right to 24 hours.
Upper panels: 30 GHz channel; lower panels: 100 GHz channel.
}
\label{fig:alpha_det}
\end{figure}

The ``weight function" technique allows us to improve the calibration accuracy.
Results on the optimization of $\alpha$ are shown in Fig. \ref{fig:alpha_det}.
The optimum choice of $\alpha$ depends on both frequency 
and integration time. The frequency dependency is obvious, since 
foreground emission and noise levels are different at the various 
frequencies.
The integration time dependence comes from the fact that
the systematic error is independent of time, while white noise 
scales as $1/\sqrt{\tau}$.
The $\alpha$ step in simulations is 0.125 at both frequencies.
On the 1-hour time scale, we find that the best result is with 
$\alpha=1.5$ at 30 GHz and 
$\alpha=0.625$ at 100 GHz; 
on a 24-hour time scale the best result is with 
$\alpha=2.875$ at 30 GHz and 
$\alpha=0.75$ at 100 GHz.

\begin{figure}[htb]
\mbox{}
\centerline{\epsfig{figure=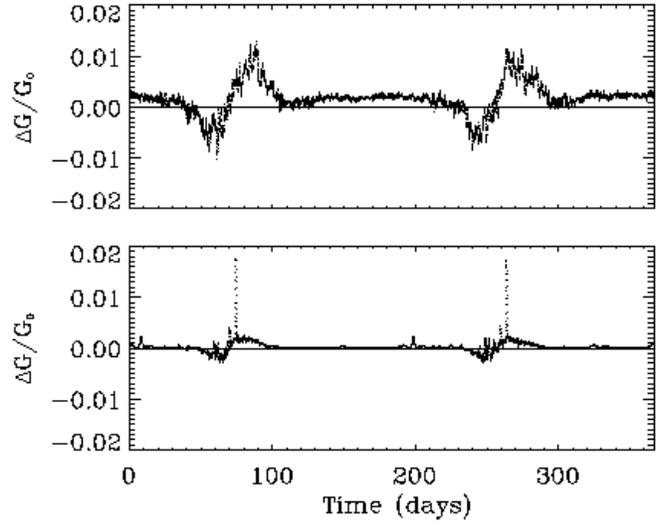, width=10cm, angle=0}}
\caption[]{Systematic error on $G$ calculated every hour, 
with $\alpha=1.5$ at 30 GHz (upper panel) and 
with $\alpha=0.625$ at 100 GHz (lower panel).}
\label{fig:systpot_1h}
\end{figure}

\begin{figure}[htb]
\mbox{}
\centerline{\epsfig{figure=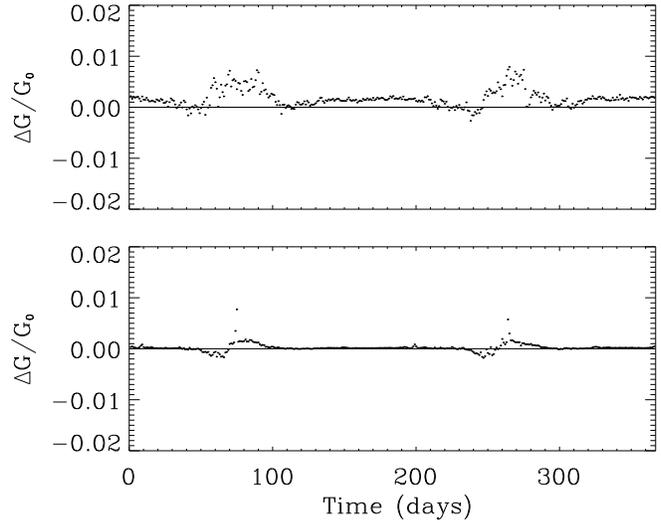, width=10cm, angle=0}}
\caption[]{Systematic error on $G$ calculated every 24 hours, 
with $\alpha=2.875$ at 30 GHz  (upper panel) and 
with $\alpha=0.75$ at 100 GHz (lower panel).}
\label{fig:systpot_24h}
\end{figure}

Results for systematic errors are shown in Figs. \ref{fig:systpot_1h} and 
\ref{fig:systpot_24h}. 
Considering an integration time of 1 day, one can achieve $\sigma_{syst}$ 
$\le 1\%$ for the whole mission time at 30 GHz
(Fig. \ref{fig:systpot_24h}, upper panel) and
$\sigma_{syst}$ is $\le 0.3\%$ pratically for $100\%$ of the mission time 
at 100 GHz (Fig. \ref{fig:systpot_24h}, lower panel).
These results are in line with the requirements indicated in Appendix 
\ref{app:requirements}.

\section{CMB Anisotropies}
\label{sect:cmb}

Finally we considered the impact of the presence of CMB intrinsic
anisotropies on the calibration accuracy.
Anisotropy maps were simulated with the {\tt SYNFAST} 
routine of the HEALPix package, 
given a Standard Cold Dark Matter CMB power spectrum.
The spectrum was obtained through the CMBFAST 
code\footnote{\tt http://physics.nyu.edu/matiasz/CMBFAST/cmbfast.html}.
We used the best calibration parameters found in our foregrounds 
analysis.

The same anisotropy realizations, properly converted into antenna temperature,
were used at both frequencies; we then obtain the observed sky: 
\beq
T^{\rm obs}=T^{\rm dip}+T^{\rm gal}+T^{\rm cmb} 
\enq
and the calibration sky:
\beq
T^{\rm cal}=T^{\rm dip}+T^{\rm gal} \, .
\enq
This is equivalent to neglecting CMB anisotropies in the calibration procedure:
we assume no information at all on their presence.

\begin{figure}[h!]
\vspace*{-0.6cm}
\hspace*{0.2cm}
\mbox{}
\centerline{\epsfig{figure=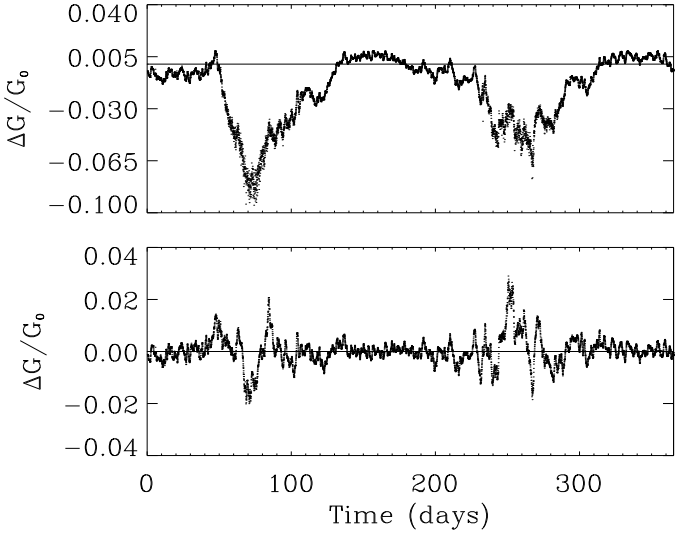, width=10cm, angle=0}}
\vspace*{-1.5cm}
\caption[]{30 GHz.
Upper panel: systematic error on calibration due to CMB anisotropies. 
Lower panel: systematic error calculated taking into account information 
we have on CMB amplitudes on large ($\theta \ge 7^{o}$) scales.}
\label{fig:cmb30}
\end{figure}

\begin{figure}[h!]
\vspace*{-0.6cm}
\hspace*{0.2cm}
\mbox{}
\centerline{\epsfig{figure=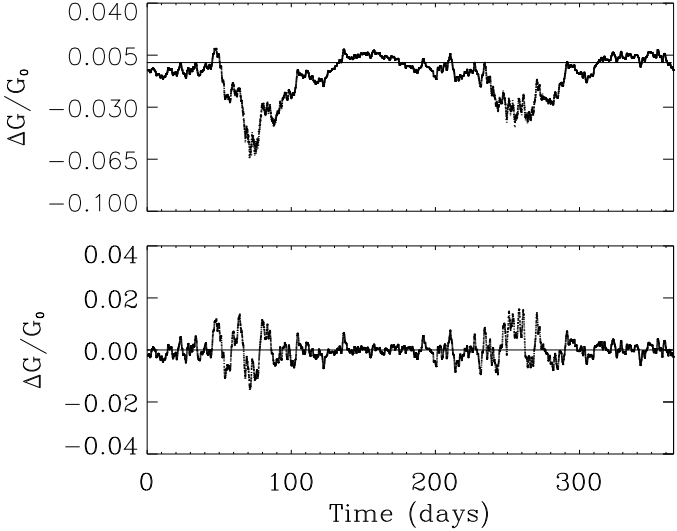, width=10cm, angle=0}}
\vspace*{-1.5cm}
\caption[]{100 GHz.
Upper panel: systematic error on calibration due to CMB anisotropies. 
Lower panel: systematic error calculated taking into account information 
we have on CMB amplitudes on large ($\theta \ge 7^{o}$) scales.}
\label{fig:cmb100}
\end{figure}

Results on systematic errors are shown in the upper panels of 
Figs. \ref{fig:cmb30} and \ref{fig:cmb100}. Note that the deviations are 
quite significant, exceeding $5\%$ in some periods of the observations.
The largest deviations are due to large scale 
CMB anisotropy structures, as one can anticipate based on the
{\sc Planck} scanning strategy. In fact,
if we consider the information on CMB 
amplitudes on large scales (such as the $COBE$-DMR ones 
on $\theta \geq 7^{\circ}$) in the calibrator sky, 
and LFI-like angular resolution CMB map in the observed sky, 
so that 
\beq
T^{\rm obs}=T^{\rm dip}+T^{\rm gal}+T^{\rm cmb}_{\rm LFI-FWHM}
\enq
\beq
T^{\rm cal}=T^{\rm dip}+T^{\rm gal}+T^{\rm cmb}_{\rm COBE-FWHM} \, \, ,
\enq
only small scale structures continue to impact calibration, giving an error 
``randomly" distributed around zero (Fig. \ref{fig:cmb30} and \ref{fig:cmb100}, lower 
panels). To maintain the systematic error of calibration within $\sim 1\%$, 
it is therefore necessary to use the information on the actual distribution 
of cold and hot spots in the CMB at large scales (e.g. $\gsim 7^{\circ}$).

\section{Conclusions}
\label{sect:conclusions}

The CMB dipole provides a nearly ideal source for calibration of extended 
maps of CMB anisotropy. 
For sensitive, full-sky surveys the modulation of
the CMB dipole due to Earth rotation provides the most precisely known
absolute calibration signal. However, preliminary absolute calibration on
short-time scales needs to rely on the observed sky signal. Also,
experiments mapping limited sky regions need to use the microwave sky as the
calibration source. In this paper we have discussed an optimisation
strategy using the sky emission as an absolute calibration source.
In the absence of galactic foreground contributions, 
which are currently known with rather poor precision at millimeter 
wavelengths, high sensitivity, high resolution experiments 
could achieve calibrations accurate at $\leq$ 1$\%$ level 
on a time scale of 1 hour for nearly the entire mission. 
However, in practice, the presence of galactic diffuse 
emission introduces important systematic effects due to the  
uncertainties on their intensity and spatial structure. Cuts in galactic 
latitude can be used to mitigate this problem. However, 
residual systematic deviations in the recovery of the calibration gain are 
significant (typically  at the $10\%$ level) due to the presence 
of high galactic latitude structures.
We have shown that introducing a suitable weight function based on the estimated 
uncertainty in the total signal per pixel, it is possible to improve 
significantly the calibration accuracy with respect to methods simply involving 
cuts in galactic latitude. 
We applied this concept to the case of {\sc Planck-LFI}, and produced 
quantitative estimates of the calibration accuracy with dedicated simulations.
We have computed the optimal parameter $\alpha$ for the weight function, 
which, for a given experiment, depends on the frequency and time 
scale of calibration.

The results show that {\sc LFI} with its nominal scanning strategy
can reconstruct every hour the value of the gain parameter 
with 1\% accuracy for 98\% of the total time of the mission at 30 GHz and 
for 99.5\% at 100 GHz. 
If we consider an integration time of 1 day we obtain an accuracy 
$\le 1\%$ for 100\% of the time at 30 and 100 GHz.
Finally we have shown that the impact of the presence of the CMB anisotropies 
themselves is less than $1-2\%$ for most of the time 
using the knowledge of their large-scale 
distribution from $COBE$-DMR.

While these results have been obtained for the {\sc Planck-LFI} survey, the 
basic method can be applied to any precision measurements of the CMB on 
large sky areas.
The recent measurements at mm wavelenghts by WMAP help to improve 
the a priori knowledge of the galactic signal. 
These new results can be readily incorporated in the proposed technique, 
leading to a more accurate determination of the  
optimum values of the $\alpha$ parameter; we will consider this 
in a forthcoming paper.

\vspace*{0.2cm}
Future works include the analysis of the impact of instrumental 
systematics on the calibration procedure; we also need to study
their behaviour with respect to the weight-function technique proposed
in this paper.


\appendix
\section{Calibration requirement}
\label{app:requirements}

The required precision on calibration depends on the goal precision  
in recovering the angular power spectrum coeficients $C_{\ell}$,
typically a few \% for precision experiments; 
this ensures cosmological parameters with comparable accuracy.
We derive here a relation between a systematic error on the calibration
-the $G$ factor- and the accuracy on the recovered power spectrum $C_\ell$,
in the simple case where $G$ is constant over the whole map.

An error $\delta G$ on the gain factor induces an error on 
the measured $\Delta T$:
\begin{eqnarray}
\Delta T^{'} & = &  G^{'} \Delta V = 
                    G \left( 1+\frac{\delta G}{G} \right) \Delta V = \nonumber \\
             & = & \left( 1+\frac{\delta G}{G} \right) \Delta T
\label{a2}
\end{eqnarray}
then
\beq
\label{a3}
\Delta T^{'}-\Delta T = \frac{\delta G}{G} \Delta T.
\enq
In the first case the coefficients of the spherical decomposition of the 
$\Delta T$ function are
\beq
        a_{\ell m}=\int \frac{\Delta T}{T}(\theta,\phi) 
        Y_{\ell m}(\theta,\phi)d\,\Omega \, ,
\enq
while in the second one
\beq
\label{a4}
        a_{\ell m}^{'}=\int \frac{\Delta T^{'}}{T}(\theta,\phi)
        Y_{\ell m}(\theta,\phi)d\,\Omega \, .
\enq
Subtracting the previous two expression,
\begin{eqnarray}
\label{a5}
        \delta a_{\ell m}&=&\int \frac{\Delta T^{'}-\Delta T}{T}(\theta,\phi)
        Y_{\ell m}(\theta,\phi)d\,\Omega \nonumber \\
        &=& \int \frac{\delta G}{G} \frac{\Delta T}{T} (\theta,\phi)
        Y_{\ell m}(\theta,\phi)d\,\Omega  \nonumber \\
        &=& \frac{\delta G}{G}a_{\ell m} \, . 
\end{eqnarray}
We recall the relation between $C_{\ell}$ and $a_{\ell m}$ coefficients:
\beq
         C_{\ell}=\frac{1}{2\ell +1}\sum_{m=-\ell}^{\ell}| a_{\ell m}|^2 \, ;
\enq
differentiating
\begin{eqnarray}
\delta C_{\ell}&=&\frac{\partial C_{\ell}}{\partial a_{\ell m}} \delta a_{\ell 
m} = 
\frac{2}{2\ell +1}\sum_{m=-\ell}^{\ell}| a_{\ell m}| \delta a_{\ell m} 
\nonumber \\
&=&\frac{2}{2\ell +1} \frac{\delta G}{G} \sum_{m=-\ell}^{\ell}| a_{\ell m}|^2 
= 
2 \frac{\delta G}{G} C_{\ell}
\end{eqnarray}

\beq
\rightarrow \frac{\delta C_{\ell}}{C_{\ell}} = 2 \frac{\delta G}{G} \, .
\enq
In conclusion, in order to get $\delta C_{\ell} / C_{\ell} \simeq 1-2 \%$ 
one requires 
\beq
\frac{\delta G}{G} \leq 1 \% \, .
\label{Grequirement}
\enq

\end{document}